\documentclass[3p,times,twocolumn]{elsarticle}

\usepackage{ecrc}

\volume{00}

\firstpage{1}

\journalname{Nuclear Physics B Proceedings Supplement}

\runauth{}

\jid{nuphbp}

\jnltitlelogo{Nuclear Physics B Proceedings Supplement}

\usepackage{amssymb}

\usepackage{lineno}

\usepackage[figuresright]{rotating}

\begin{document}
\begin{frontmatter}



\dochead{}

\title{Data processing and storage in the Daya Bay Reactor Antineutrino Experiment}


\author{Miao He}
\ead{hem@ihep.ac.cn}

\address{Institute of High Energy Physics, Beijing}
\address{On behalf of the Daya Bay collaboration}

\begin{abstract}
The Daya Bay Reactor Antineutrino Experiment reported the first observation of the non-zero neutrino mixing angle $\theta_{13}$ using the first 55 days of data. It has also provided the most precise measurement of $\theta_{13}$ with the extended data to 621 days. Daya Bay will keep running for another 3 years or so. There is about 100~TB raw data produced per year, as well as several copies of reconstruction data with similar volume to the raw data for each copy. The raw data is transferred to Daya Bay onsite and two offsite clusters: IHEP in Beijing and LBNL in California, with a short latency. There is quasi-real-time data processing at both onsite and offsite clusters, for the purpose of data quality monitoring, detector calibration and preliminary data analyses. The physics data production took place a couple of times per year according to the physics analysis plan. This paper will introduce the data movement and storage, data processing and monitoring, and the automation of the calibration.
\end{abstract}


\end{frontmatter}


\section{Introduction}
Neutrinos are elementary particles in the Standard Model.
There are three flavors of neutrinos, known as $\nu_{e}$, $\nu_{\mu}$ and $\nu_{\tau}$.
Neutrinos exist in the nuclear fusion in the sun,
$\beta$-decays of radioactivities inside the Earth,
the Big Bang remnant, the supernova explosion, or interactions
between cosmic rays and the atmosphere of the Earth.
Besides the natural sources, neutrinos can be also produced artificially in reactors and accelerators.
The neutrino flavor states are superpositions of three mass eigenstates ($\nu_{1}$, $\nu_{2}$ and $\nu_{3}$).
One flavor can change to another because of the
quantum interference of the mass eigenstates during the traveling.
This phenomenon is known as neutrino oscillation or neutrino mixing.
The amplitude of the oscillation is connected to the mixing angles $\theta_{12}$, $\theta_{23}$ and $\theta_{13}$.
The oscillation frequencies are determined by the difference
of squared neutrino masses, ${\Delta}m^{2}_{ij} = m^{2}_{i} - m^{2}_{j}$.
For reactor-based experiments, $\theta_{13}$ can be extracted from the survival probability
of the electron antineutrino at a distance of 1$\sim$2 km from the reactors
\begin{equation}
P(\overline{\nu}_{e}\rightarrow\overline{\nu}_{e}){\approx}1-\sin^{2}2\theta_{13}\sin^{2}(1.27{\Delta}m^{2}_{31}L/E)
\end{equation}
\noindent where $E$ is the $\overline{\nu}_{e}$ energy in MeV and $L$ is the distance
in meters between the $\overline{\nu}_{e}$ source and the detector (baseline).

\section{The Daya Bay experiment}
\begin{figure}[htb]
\center{\includegraphics[width=\columnwidth]{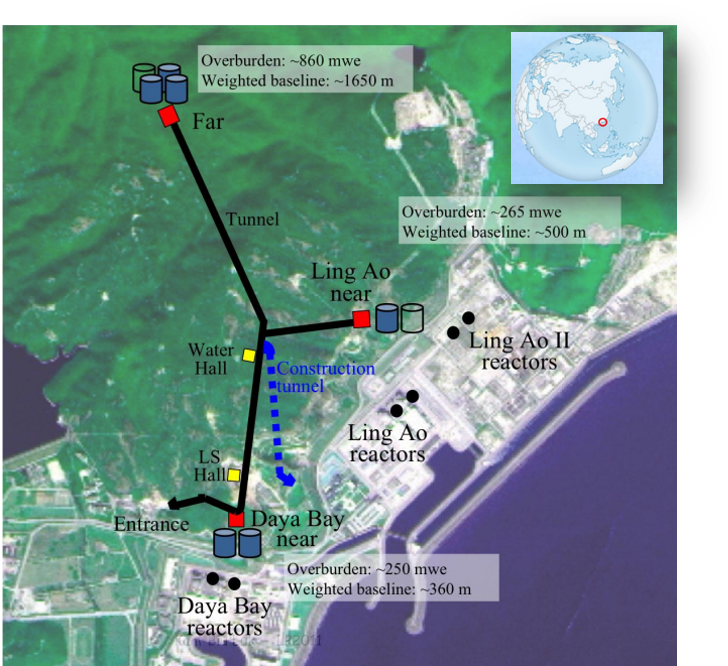}}
\caption{Layout of the Daya Bay experiment.}
\label{dyblayout}
\end{figure}

The Daya Bay experiment was designed to provide the most precise
measurement of $\theta_{13}$ among existing and near future experiments,
with a sensitivity to $\sin^{2}2\theta_{13} < 0.01$ at the 90\% CL~\cite{Guo:2007ug}.
The experiment is located near Daya Bay, Ling Ao and Ling Ao-II nuclear power plants in southern China,
45 km from Shenzhen city. As shown in Fig.~\ref{dyblayout},
there are six functionally identical reactor cores, grouped into three pairs.
Three underground experimental halls (EHs) are connected with horizontal tunnels.
Two antineutrino detectors (ADs) were located in the Daya Bay near hall (EH1),
two in the Ling Ao and Ling Ao-II near hall (EH2),
and four near the oscillation maximum in the far hall (EH3).
Each AD is filled with 20 t of gadolinium-doped liquid scintillator (Gd-LS) as a target,
surrounded by 20 t un-doped liquid scintillator (LS), for detecting gammas that escape from the target.
The $\overline{\nu}_{e}$ interacts with proton via the inverse $\beta$-decay (IBD) reaction
in Gd-LS, and releases a $e^{+}$ and a neutron. The prompt signal from $e^{+}$ ionization
and the delayed signal from neutron capture on Gd with $\sim$30 $\mu$s latency
provides a time coincidence signature.
The muon veto system in each hall consists of an inner water shield, an outer water shield
and a resistive plate chamber (RPC).
The detailed introduction to the Daya Bay experiment can be found in Ref.~\cite{DayaBay:2012aa}.

Daya Bay started data taking in the end of 2011 with six ADs.
The last two ADs were installed in summer 2012.
Multiple partitions in the data acquisition software (DAQ)~\cite{Li:2011am}
allow data taking in three experiment halls
simultaneously and the output of multiply data streams.
A typical physics run in each hall lasts for 48 hours, with
a pedestal run and an electronics diagnosis run in between.
The typical trigger rates are 1.3~kHz, 1.0~kHz, 0.6~kHz in EH1, EH2, EH3, respectively.
The total number of raw data files each day is about 320,
approximately 1 GB volume per file.
The ratio of the data taking time in 2013 is larger than 97\%,
and of the physics data taking time is larger than 95\%.
Based on the first 55 days of data, Daya Bay reported the first observation
of the non-zero neutrino mixing angle $\theta_{13}$ with 5.2 standard deviations~\cite{An:2012eh}.
The precision of $\sin^{2}2\theta_{13}$ was continuously updated with increased statistics.
The latest result of Daya Bay was based on 621 days of data with a precision of 5.6\%~\cite{DB8AD}.

\section{Data processing and storage in Daya Bay}

\begin{figure}[htb]
\center{\includegraphics[width=\columnwidth]{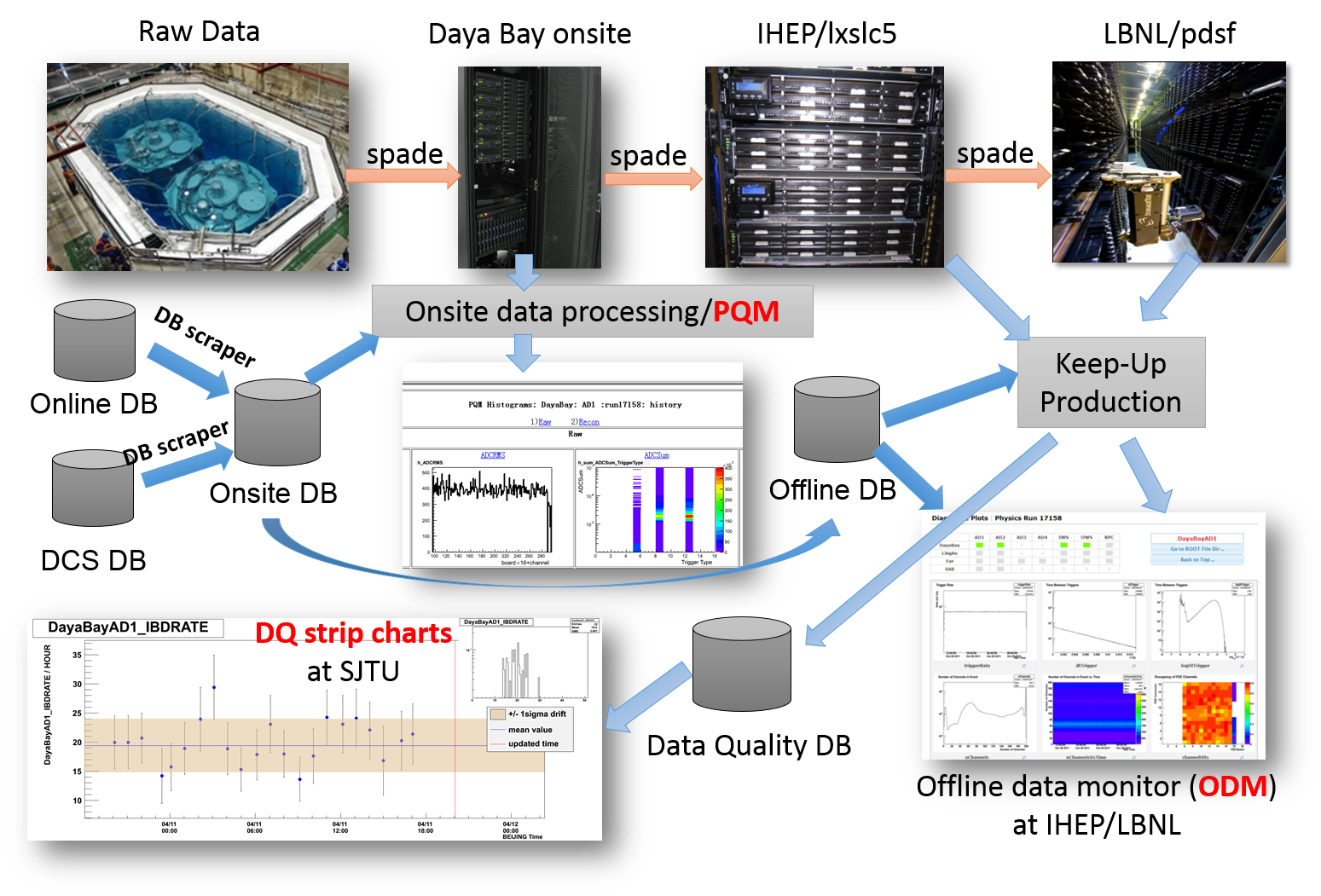}}
\caption{Overview of the data processing in Daya Bay.}
\label{data_processing}
\end{figure}

Fig.~\ref{data_processing} is a global picture of Daya Bay data processing.
Raw data is firstly transferred to the onsite farm then to
Insititute of High Energy Physics (IHEP) in Beijing and
Lawrence Berkeley National Laboratory (LBNL) in California as central storage and processing.
The metadata in the online database is also transferred to the offline database.
A Performance Quality Monitoring system (PQM)~\cite{Liu:2014rpa} is running onsite
using fast reconstruction algorithms to monitor the physics performance
with a latency of around 40 minutes.
A keep-up data processing takes place as soon as the data reach IHEP or LBNL,
using the full reconstruction and the latest calibration constants.
The generated detector monitoring plots are published through
an Offline Data Monitoring system (ODM) with a latency of around 3 hours.
The extracted data quality information is filled in a dedicated database for the long-term monitoring.

\subsection{Data movement and storage}
The data movement is controlled by a set of Java applications, and the primary application is call SPADE.
A raw data file is copied to an offine file server at Daya Bay onsite automatically
once it's closed and recorded in the online database by DAQ.
When the movement is done, the file status in both of the online and offline database
is changed to TRANSFERRED by SPADE
so that this file can be deleted from the online disk.
After that, the file is transferred to two computing centers
IHEP and LBNL sequentially with a maximum latency of 20 minutes.
SPADE also creates a xml file which contains descriptions of raw data for the file cataloging.
Multiply tools are deployed to monitor the network traffic at Daya Bay onsite
and to monitor the rate of transferred raw data file number and data volume
from Daya Bay to IHEP, and from IHEP to LBNL.

Raw data is stored at Daya Bay onsite for one month and archived at IHEP and LBNL,
with one copy on disk and two copies on tape for each of them.
The infrastructure of these two clusters are different but the capability of computing and storage are similar.
Each cluster has near 1~PB disk space and around 800 CPU cores including some shared resources.
Additional computing resources have been planned to accommodate increasing data.
Accumulated raw data is 320~TB by the middle of 2014.

\subsection{Offline software}
 The Gaudi~\cite{gaudi} framework based Daya Bay offline software (known as NuWa) provides the full functionality required by simulation, reconstruction and physics analysis. NuWa employs Gaudi's event data service as the data manager. Raw data, as well as other offline data objects, can be accessed from the Transient Event Store (TES). The prompt-delayed coincidence analysis requires looking-back in time, which is fulfilled via specific implemented Archive Event Store (AES). All the data objects in both TES and AES can be written into or read back from ROOT~\cite{root} files through various Gaudi converters. An alternative Lightweight Analysis Framework (LAF) was designed and implemented by Daya Bay to improve the analysis efficiency. LAF is compatible with NuWa data objects with higher I/O performance by the simpler data conversion, the implementation of lazy loading, and the flexible cycling mechanism. LAF allows to access events both backwards and forwards through an data buffer, which also serves to exchange data among multiple analysis modules. The NuWa and LAF packages are available to collaborators using Subversion (SVN) code management system~\cite{svn}. The NuWa auto-build system was implemented using
the Bitten~\cite{bitten} plug-in for trac~\cite{trac}, which is an enhanced
wiki and issue tracking system for software development projects.
Multiple servers at Daya Bay onsite, IHEP, LBNL and other institutes work as the
Bitten slave and build NuWa automatically when the code is updated.

\subsection{Database}
\begin{figure}[htb]
\center{\includegraphics[width=\columnwidth]{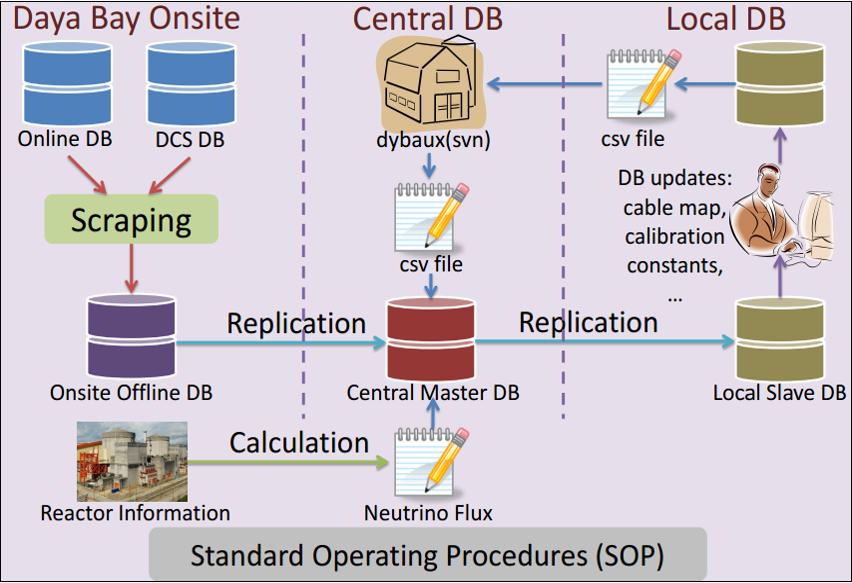}}
\caption{Standard operating procedures of the Daya Bay offline database.}
\label{database}
\end{figure}

The offline database consists of an onsite offline database, the central master database and multiple local slave databases,
as shown in Fig.~\ref{database}.
During data taking, the information such as the description of raw data and the detector status is extracted from online databases
to the onsite offline database automatically with a scraper.
Then the onsite offline database is synchronized to the central master database at IHEP.
The central database also contains the calculated reactor neutrino flux
based on the reactor information provided by the power company.
There are multiple slaves as replications of the central database located at different institutes.
Offline calibration constants produced by calibration experts are firstly filled in a temporary table of a local slave,
then dumped into a formative text file after validation.
The text file is archived in the subversion system.
The database manager is responsible to fill the calibration constants into the central database
through the text file with a revision number specified by the calibration expert.
This standard operating procedure makes sure the updating of calibration constants
highly controlled, carefully recorded and easily reproducible.

\subsection{Calibration and reconstruction}
Various of sources are used to calibrate the detector response.
Specific calibrations runs are taken weekly using LED and radioactive sources.
Other calibration samples such as PMT dark noises and neutrons produced by cosmic rays
are selected in normal physics runs.
There are two categories of the calibration.
The file-by-file track automatically accumulates calibration samples from physics runs
and generates calibration constants.
The run-by-run track uses a script to search all raw data from calibration runs
and submit jobs to get calibration constants.
Reconstruction algorithms read calibration constants from the local slave database through a Database Interface (DBI).
A timestamp called rollback date is set to choose the latest records
that were inserted into the offline database before it, for the purpose of the version control.

\subsection{Onsite data processing and monitoring}
\begin{figure*}[htb]
\center{\includegraphics[width=14cm]{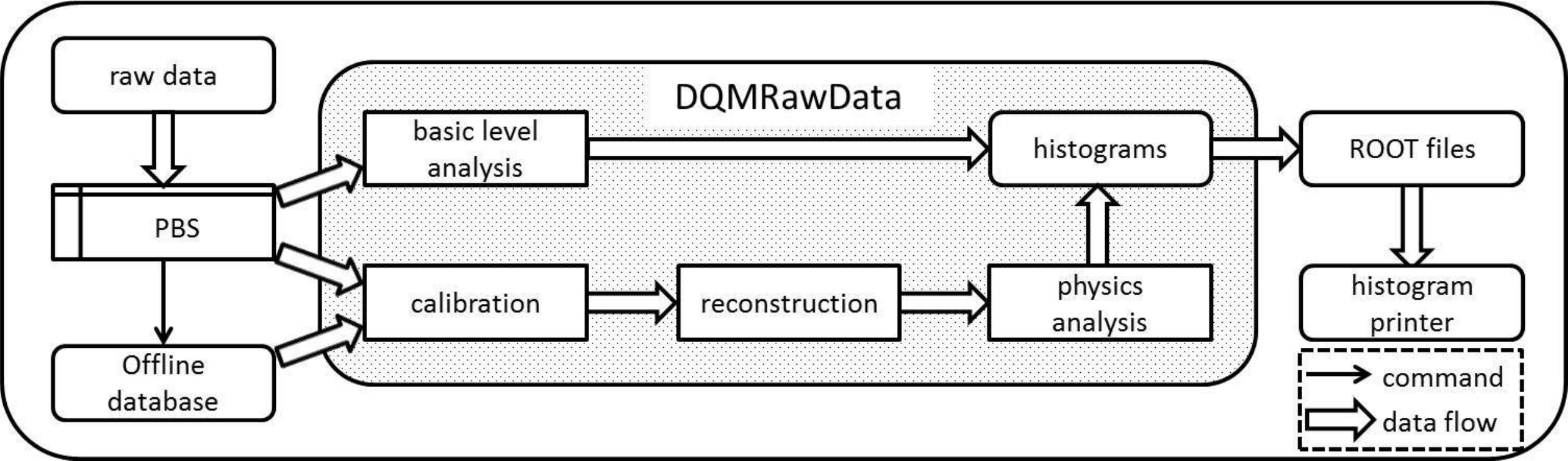}}
\caption{The data flow of the PQM job.}
\label{pqm}
\end{figure*}
The PQM system running at Daya Bay onsite
provides a quasi realtime data processing and monitoring during data taking.
There are 16 dedicated CPU cores for PQM and 40 additional cores shared with users.
A Python control script has been developed and
runs in background mode on a dedicated server,
which queries the onsite offline database at a
fixed time interval (10 seconds). If a new record is found
and the data status returned by the query shows that the
corresponding raw data file is already on the file server, a
job for processing the new file is submitted to the Portable Batch System (PBS).
The PQM job uses a recently manually tagged NuWa version
to reconstruct data with the latest calibration constants
in the offline database, runs analysis algorithms,
and fills the results into user-defined histograms. All of
the histograms are dumped into a ROOT file at the
finalization step of the algorithms, which is then merged
with the accumulated ROOT file for the same run. After
that, a C++ histogram printer is executed, taking the
merged ROOT file as input to print selected histograms
into figures. The control script also checks if the job
is finished at 10-second intervals by detecting an empty
TXT file created at the end of the job. For finished jobs,
the control script transfers the corresponding figures to
the PQM disk for web display. When the
run ends and all the corresponding raw data files are processed,
the control script saves the accumulated ROOT
file on the PQM disk for permanent storage.
The data flow of PQM in shown in Fig.~\ref{pqm}.
The reconstruction and analysis modules in PQM is configurable,
and the default configuration takes around 30~minutes for each job.

\subsection{Offsite data processing and monitoring}
\begin{figure}[htb]
\center{
\includegraphics[width=\columnwidth]{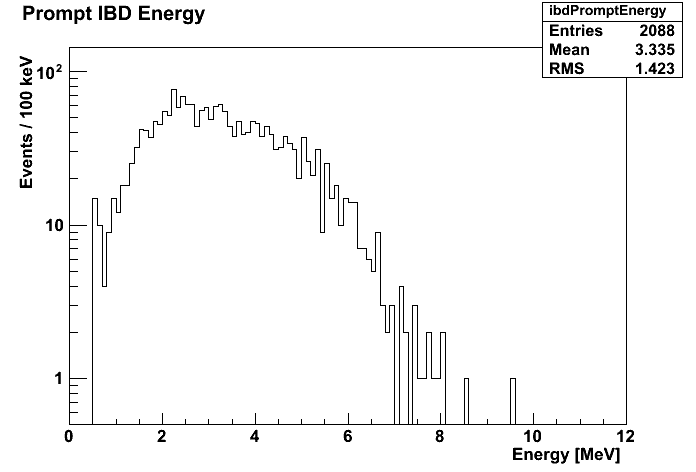}
\includegraphics[width=\columnwidth]{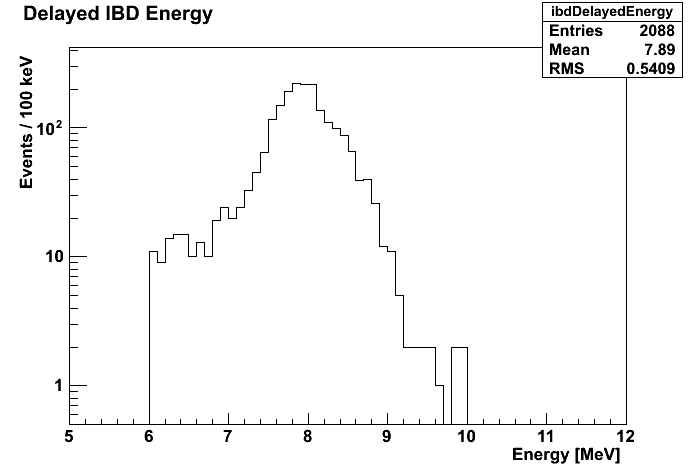}
}
\caption{Monitoring of the energy spectra of prompt (up)
and delayed (down) signals from the inverse $\beta$-decay on ODM.}
\label{odm}
\end{figure}

As soon as a raw data file reaches IHEP or LBNL,
an application called ingest in the data movement system
submits a job using the full NuWa reconstruction and the latest calibration constants,
known as the keep-up (KUP) data processing.
More analysis modules are running allowing a high-level monitoring of data.
The output histograms categorized by different detectors and by different channels
are archived in ROOT files and printed as figures.
A multifunctional webpage has been developed in the framework of Django,
named as the Offline Data Monitor (ODM).
The major function of ODM is to display the plots produced in KUP.
It provides additional tools to help the data monitoring and analysis,
such as the comparison to references, the display of detector configuration and status,
and interfaces of online and offline databases.
Two example plots on ODM are shown in Fig.~\ref{odm}.
The prompt and delayed signals from the inverse $\beta$-decay are selected based on AES during KUP.
The KUP jobs have the highest priority in the job system,
which typically occupy 30 CPU cores.
The latency of ODM is dominated by the reconstruction and analysis algorithms.

Physics production uses the validated and frozen calibration constants and reconstruction algorithms.
It also contains event tagging and filtering and provides both the full data sample and different reduced samples
to improve the efficiency of data analysis.
The reconstructed data volume is about 1.2 times of raw data, and only the latest version
is kept on disk, while previous versions that used for the publication were archived on tape.
Physics production takes place one or two times per year according to the requirement of the physics analysis,
with about one month of processing time for each production.
A special production strategy was used for the first two publications~\cite{An:2012eh,An:2013uza} to quickly get the physics result.
The offline software was fixed in the beginning. The calibration constants were generated after
the weekly calibration data taking, followed by the extension of the physics data production for the past week.
Thanks to this weekly production strategy, Daya Bay finished the analysis and reported the first physics result
only 20 days after the first stage of data taking.

\subsection{Data quality}
\begin{figure}[htb]
\center{\includegraphics[width=\columnwidth]{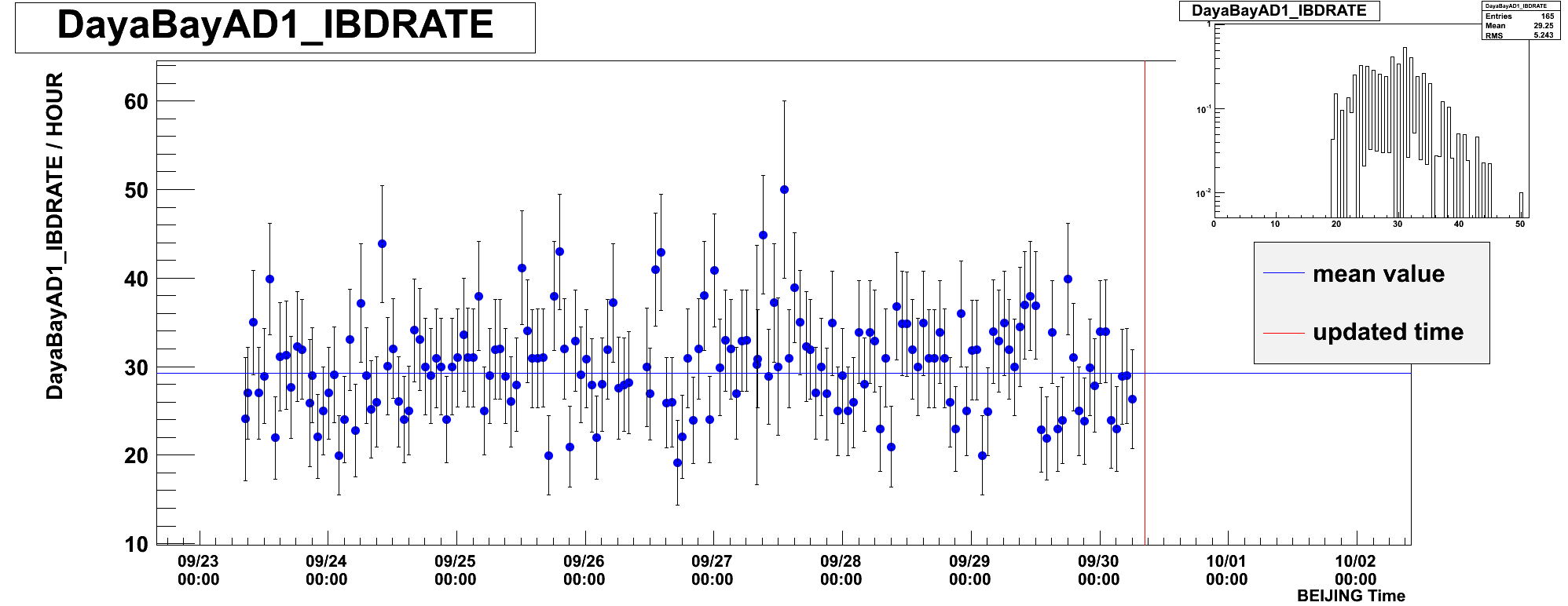}}
\caption{Monitoring of the antineutrino candidate rate.}
\label{dq}
\end{figure}

During the keep-up data processing, information related to the data quality
is extracted and filled in a dedicated database,
which is automatically checked by an application for the evaluation of the data quality.
On the other hand, online shifters record any issue of data in the database.
A data quality manager is responsible to make a tag to each raw data file
based on both auto-check and manual-check results,
and provides a preliminary good file list.
After the physics data production, the final good file list which is used
for the physics publication is compiled based on the feedback from analyzers.
A web interface of the data quality database is implemented for the long-term monitoring of the data quality.
An example is Fig.~\ref{dq} which shows a strip chart of the antineutrino candidate rate
in one of the AD in EH1.

\section{Summary}
Daya Bay is a reactor antineutrino oscillation experiment.
It provided the first and most precision measurement of the neutrino mixing angle $\theta_{13}$.
The offline system provided quick and stable data transferring, data processing and data quality monitoring.
Data Bay will continue taking data till the end of 2017. The computing resources and the offline software
will be upgraded to accommodate the data processing and data storage.




\nocite{*}
\bibliographystyle{elsarticle-num}
\bibliography{ref}







\end{document}